\documentclass{elsart}
\usepackage{graphics}

\begin{document}

\begin{frontmatter}
\title{Influence of rare regions on quantum phase transition 
in antiferromagnets with hidden degrees of freedom}
\author{Y. N. Skryabin\thanksref{RFBR}}
\address{Institute for Metal Physics, Russian Academy of Sciences,\\
Ural Division, Kovalevskaya Str., 18,\\
620219, Ekaterinburg, Russian Federation \\}
\author{A. V. Chukin \and A. V. Shchanov}
\address{Ural State Technical University, \\
Mir Str., 19,
620002, Ekaterinburg, Russian Federation}
\thanks[RFBR]{Partially supported by Russian Foundation for Basic 
Research (97-02-17315), Russian Federation.}

\begin{abstract}
The effects of rare regions on the critical properties   
of quantum antiferromagnets with hidden degrees of freedom within the 
renormalization group is discussed. It is shown that for 
``constrained'' systems the stability range on the phase diagram 
remains the same as in the mean-field theory while for ``unconstrained'' 
systems the stability range is effectively decreased.
\end{abstract}
\end{frontmatter}

\section{Introduction}

The influence of quenched disorder on the critical properties 
of itinerant quantum magnets is a very interesting problem 
in phase transition theory. 
Particular interest was given to locally ordered spatial 
regions (``rare regions'') which are formed in the presence 
of quenched disorder even when the bulk system is still in 
the paramagnetic phase \cite{Griffiths}. 
The conventional theory \cite{Harris} ignores these rare 
regions.   
For classical magnets, it has recently been shown that the 
rare regions induce a new term in the action which breaks 
the replica symmetry and that the conventional theory is 
unstable with respect to this perturbation \cite{Dotsenko}. 

The problem of rare regions for the case of quantum phase 
transitions has been considered in 
Refs.\ \cite{Narayanan99a,Narayanan99b}.  
These authors showed that the rare regions destroy the fixed 
point found in the conventional theory of itinerant 
antiferromagnets \cite{Kirkpatrick96a} and, in contrast, 
critical behavior of itinerant ferromagnets 
\cite{Kirkpatrick96b} is unaffected by the rare regions due 
to an effective long-range interaction between the order 
parameter fluctuations. 

The conventional theory of a random-$T_c$ quantum 
antiferromagnet with the hidden degrees of freedom on
which various constraints are imposed has been considered 
in Ref.\ \cite{Skryabin99b} and it has been shown that for 
``constrained'' systems the stability range on the phase 
diagram remains the same as in the mean-field theory while 
for ``unconstrained'' systems the stability range 
is effectively decreased.
In this paper we study the effects of rare regions on the 
quantum phase transition in antiferromagnets with hidden 
degrees of freedom. 

\section{Renormalization group equations}

Let us consider a disordered itinerant quantum antiferromagnet
in which the $n$-component vector order parameter 
${\bf S}({\bf r}, \tau)$ is coupled with the scalar 
nonfluctuating parameter $y ({\bf r}, \tau)$. 
We use an action  
\begin{eqnarray}
S &=& \frac12 \int \d {\bf r}_1 \d {\bf r}_2 \int_0^{1/T} 
        \d \tau \d \tau ' [t({\bf r}_1 
        - {\bf r}_2, \tau - \tau ') \nonumber \\ 
& &+\delta ({\bf r}_1- {\bf r}_2) \delta (\tau - \tau ') 
      \delta t({\bf r})]
{\bf S}({\bf r}_1, \tau) {\bf S}({\bf r}_2, \tau ')\nonumber\\
& &+\lambda_s \int \d {\bf r} \int_0^{1/T} \d \tau 
[{\bf S}({\bf r}, \tau ) {\bf S}({\bf r}, \tau )]^2 \nonumber\\
& &+\mu \int \d {\bf r} \int_0^{1/T} \d \tau y({\bf r}, \tau)
[{\bf S}({\bf r}, \tau)]^2\nonumber\\
& &+\frac12 \beta \int \d {\bf r} \int_0^{1/T} \d \tau  
[y({\bf r}, \tau)]^2.
\label{action}
\end{eqnarray}
Here the function $t({\bf r}, \tau)$ is the Fourier-transform 
of the two-point interaction of a quantum antiferromagnet 
\cite{Hertz,Kirkpatrick96a}
\begin{equation}
t({\bf q}, \omega_n) = t + {\bf q}^2 + \omega_n,
\label{t_qn}
\end{equation}
where $t$ denotes the distance from the quantum critical 
point, ${\bf q}$ is the wave vector and $\omega_n$ is the 
Matsubara frequency.

As the given model is phenomenological, we suppose that the 
coupling constant of the order parameter with the nonfluctuating 
parameter is purely imaginary as it occurs in the Hubbard 
model \cite{Chaves,Lyra,Skryabin97}.
It is easy to consider the case with a real coupling constant.
The distinction between these two cases consists only in the 
definition of the part of an appropriate phase space which 
corresponds to a real physical model.
In contrast to the classical phase transition the order parameter 
depends on both the $d$-dimensional vector of space $\bf r$ 
and imaginary time $\tau$ \cite{Hertz}.

In order to write an action of the conventional theory  
the replica trick should be used and then an integration over 
random variables should be performed. 
For simplicity we consider the case when the coefficient 
of the quadratic term in the order parameter in the action, 
$\delta t({\bf r})$, is the fluctuating Gaussian variable with 
zero mean and variance $\Delta$. 
Finally we obtain an action which has homogeneous saddle 
points only \cite{Skryabin99b}.

To incorporate rare regions into the theory we need a 
different approach. 
In analogy to Refs.\ \cite{Dotsenko,Narayanan99a,Narayanan99b}, 
we consider inhomogeneous saddle-point solutions for a fixed 
realization of disorder. The partition function can be 
written as the sum of all contributions obtained from the 
vicinity of each saddle-point \cite{Narayanan99b}. Then we 
can average over disorder by means of the replica trick. 
We finally obtain the following effective action 
\begin{eqnarray}
S_{\mathrm{eff}} &=& \frac12 \int \d {\bf r}_1 \d {\bf r}_2 
\int_0^{1/T} \d \tau \d \tau '
t({\bf r}_1 - {\bf r}_2, \tau - \tau ') 
\sum_{\alpha}{\bf S}^{\alpha}({\bf r}_1, \tau) 
{\bf S}^{\alpha}({\bf r}_2, \tau ')\nonumber\\
& &+ \lambda_s \int \d {\bf r} \int_0^{1/T} \d \tau \sum_{\alpha}
[{\bf S}^{\alpha}({\bf r}, \tau )
{\bf S}^{\alpha}({\bf r}, \tau )]^2 \nonumber\\
& &- \sum_{\alpha ,\beta} (\Delta + x \delta_{\alpha,\beta}) 
\int \d {\bf r} \int_0^{1/T} \d \tau \d \tau ' 
[{\bf S}^{\alpha}({\bf r}, \tau ){\bf S}^{\alpha}
({\bf r}, \tau )]\times \nonumber\\
& & [{\bf S}^{\beta}({\bf r}, \tau ' ){\bf S}^{\beta}({\bf r}, 
\tau ' )] \nonumber\\
& &+\mu \int \d {\bf r} \int_0^{1/T} \d \tau \sum_{\alpha} 
y^{\alpha}({\bf r}, \tau)
[{\bf S}^{\alpha}({\bf r}, \tau)]^2 \nonumber \\
& & +\frac12 \beta \int \d {\bf r} \int_0^{1/T} \d \tau 
\sum_{\alpha} [y^{\alpha}({\bf r}, \tau)]^2.
\label{eff_action}
\end{eqnarray}
Here $\alpha$ and $\beta$ are replica indices. The $x$-term is 
generated by taking into account the inhomogeneous saddle points. 
The conventional theory misses this term. 

Let us assume the temperature $T$ to be equal zero and use double 
$\epsilon$ expansion \cite{Kirkpatrick96a} according to which 
the space dimensionality is equal $4-\epsilon$ 
and the dimensionality of imaginary time is $\epsilon_\tau$. 
Of course, for the real physical case we have 
$\epsilon = \epsilon_\tau = 1$.

Defining $\bar x = xT^{-\epsilon_\tau}$, and putting $T=0$, 
we obtain the following renormalization group flow 
equations in the one-loop approximation    
\begin{eqnarray}
\frac{\d u}{\d l} &=& (\epsilon-2\epsilon_\tau)u - 4(n+8){u}^2 + 
                            6u\Delta, \label{u}\\
\frac{\d \Delta}{\d l} &=& \epsilon \Delta - 8(n+2)u\Delta
            + 4{\Delta}^2 + 8n\Delta \bar x, \label{delta}\\
\frac{\d \bar x}{\d l} &=& (\epsilon-2\epsilon_\tau) \bar x -  
      8(n+2)u\bar x + 4n\bar x^2 + 6\Delta \bar x,\label{x}\\
\frac{\d z}{\d l} &=& (\epsilon-2\epsilon_\tau)z - 8(n+2)uz - 
               2nz^2 + 2z\Delta,\label{z}\\
\frac{\d w}{\d l} &=& (\epsilon-2\epsilon_\tau)w - 8(n+2)uw - 
                     2nw^2 - 4nzw + 2w\Delta,\label{w}
\end{eqnarray}
where $l=\ln b$ with $b$ the scale parameter, and we have scaled 
$K_4u \rightarrow u, K_4\Delta \rightarrow \Delta, 
K_4\bar x \rightarrow \bar x, K_4z \rightarrow z, 
K_4w \rightarrow w$ with $K_4 = 1/8\pi^2$.
We also denote here
\begin{eqnarray}
u & = &\lambda_s-\frac{\mu^2}{2\beta},\\ 
z & = &\frac{\mu^2}{\beta}-\frac{\mu_0^2}{\beta_0},\\ 
w & = &\frac{\mu_0^2}{\beta_0}.
\label{notation}
\end{eqnarray}
In equations (\ref{u})-(\ref{w}) we separate the coefficient of the 
nonfluctuating parameter $y({\bf q} = 0)$ from $y({\bf q} \neq 0)$ 
because of its possible role in constraining systems 
\cite{Chaves,Achiam1}.

\section{Fixed points}

\begin{figure}
\begin{center}
\resizebox{0.5\textwidth}{!}{\includegraphics{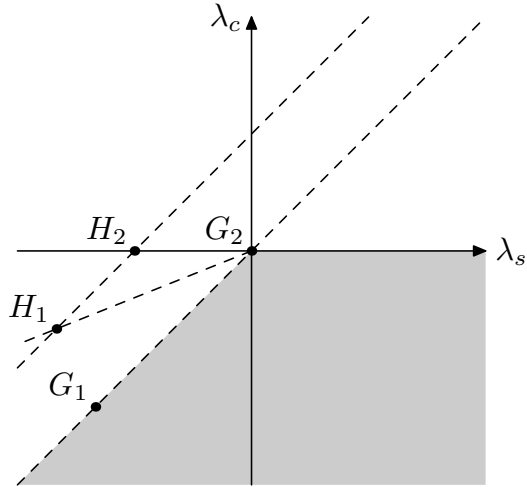}}
\end{center}
\caption{Phase diagram for the unconstrained system
($\Delta^\ast = 0, \quad \bar x^\ast = 0$)}
\label{fig:1}
\end{figure}

\begin{figure}
\begin{center}
\resizebox{0.5\textwidth}{!}{\includegraphics{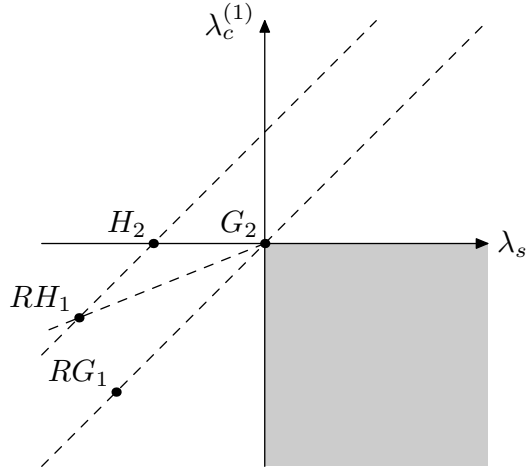}}
\end{center}
\caption{Phase diagram for the constrained system
($\Delta^\ast = 0, \quad \bar x^\ast$ = 0)}
\label{fig:2}
\end{figure}

Before discussing the renormalization group analysis we consider 
the mean-field theory result. 
After integrating over the nonfluctuating parameter in the equation 
for the partition function we can obtain a new effective 
action in terms of the order parameter ${\bf S(r)}$.
In the mean-field approximation for this effective action 
the boundaries of a stability range can be easily found. 
It is convenient to introduce new notations for the coupling constants 
of the effective action as
\begin{equation}
\lambda_c^{(0)} = \mu_0^2/2\beta_0, \quad 
\lambda_c^{(1)} = \mu^2/2\beta. 
\label{lc01}
\end{equation}
For the unconstrained system 
$\lambda_c^{(0)} = \lambda_c^{(1)} \equiv \lambda_c$ (or $z = 0$) 
while the boundaries of the stability range correspond to the 
equations of lines 
\begin{equation}
\lambda_s - \lambda_c^{(0)} = 0, \quad \lambda_c^{(0)} = 0. 
\label{stability}
\end{equation}
For the constrained system $\lambda_c^{(0)} = 0$ (or $w = 0$) 
while the boundaries of the stability range can be written as
\begin{equation}
\lambda_s = 0, \quad \lambda_c^{(1)} = 0.
\label{stabilitycon}
\end{equation}
The stability ranges are represented on planes
$(\lambda_c - \lambda_s)$ (Fig.\ \ref{fig:1}) and
$(\lambda_c^{(1)} - \lambda_s)$ (Fig.\ \ref{fig:2}).

The characteristic feature of flow equations is the closed 
system of three equations (\ref{u})-(\ref{x}).
Within the notations this set of equations coincides with the  
appropriate set of equations of Ref.\ \cite{Narayanan99a} 
where however the nonfluctuating degrees of freedom were 
not taken into account.
It is easy to find all eight fixed points (\ref{u})-(\ref{x}). 
Four of the fixed points have a zero fixed-point value of 
$\bar x^\ast$, $\bar x^\ast = 0$. 
The other four fixed points have $\bar x^\ast \neq 0$.

Using these fixed points we can find other fixed points from  
the flow equations (\ref{z}), (\ref{w}).  
There are the sixteen fixed points with $\bar x^\ast = 0$ 
(Table\ \ref{table:1}) studied before in Ref.\ \cite{Skryabin99b}. 
The other sixteen fixed points have $\bar x^\ast \neq 0$ 
(Table\ \ref{table:1a}).
\begin{table}
\caption{Fixed points with $\bar x^\ast = 0$}
\label{table:1}
\begin{tabular}{lccccc} \hline
&$u^\ast$ &$\Delta^\ast$ &$\bar x^\ast$ &$z^\ast$ &$w^\ast$ \\ 
\hline
G$_1$  & 0 & 0 & 0 & 0 &$\frac{\epsilon - 2\epsilon_\tau}{2n}$ \\
G$_2$  & 0 & 0 & 0 & 0 & 0 \\
RG$_1$ & 0 & 0 & 0 &$\frac{\epsilon - 2\epsilon_\tau}{2n}$ & 0 \\
RG$_2$ & 0 & 0 & 0 &$\frac{\epsilon - 2\epsilon_\tau}{2n}$
       &$-\frac{\epsilon - 2\epsilon_\tau}{2n}$ \\
H$_1$  &$\frac{\epsilon - 2\epsilon_\tau}{4(n+8)}$ & 0 & 0 & 0 
       &$\frac{(\epsilon - 2\epsilon_\tau)(4-n)}{2n(n+8)}$ \\
H$_2$  &$\frac{\epsilon - 2\epsilon_\tau}{4(n+8)}$ & 0 & 0 & 0 & 0 \\
RH$_1$ &$\frac{\epsilon - 2\epsilon_\tau}{4(n+8)}$ & 0 & 0 
       &$\frac{(\epsilon - 2\epsilon_\tau)(4-n)}{2n(n+8)}$ & 0 \\
RH$_2$ &$\frac{\epsilon - 2\epsilon_\tau}{4(n+8)}$ & 0 & 0 
       &$\frac{(\epsilon - 2\epsilon_\tau)(4-n)}{2n(n+8)}$  
       &$-\frac{(\epsilon - 2\epsilon_\tau)(4-n)}{2n(n+8)}$  \\
U$_1$  & 0 &$-\frac{\epsilon}{4}$ & 0 & 0 
       &$\frac{\epsilon - 4\epsilon_\tau}{4n}$ \\
U$_2$  & 0 &$-\frac{\epsilon}{4}$ & 0 & 0 & 0 \\
RU$_1$ & 0 &$-\frac{\epsilon}{4}$ & 0 
       &$\frac{\epsilon - 4\epsilon_\tau}{4n}$ & 0 \\
RU$_2$ & 0 &$-\frac{\epsilon}{4}$ & 0 
       &$\frac{\epsilon - 4\epsilon_\tau}{4n}$
       &$-\frac{\epsilon - 4\epsilon_\tau}{4n}$ \\
R$_1$  &$\frac{\epsilon + 4\epsilon_\tau}{16(n-1)}$
       &$\frac{(4-n)\epsilon + 4(n+2)\epsilon_\tau}{8(n-1)}$ & 0 
       & 0 &$\frac{(n-4)\epsilon - 12n\epsilon_\tau}{8n(n-1)}$ \\
R$_2$  &$\frac{\epsilon + 4\epsilon_\tau}{16(n-1)}$
       &$\frac{(4-n)\epsilon + 4(n+2)\epsilon_\tau}{8(n-1)}$ & 0 & 0 & 0 \\
RR$_1$ &$\frac{\epsilon + 4\epsilon_\tau}{16(n-1)}$
       &$\frac{(4-n)\epsilon + 4(n+2)\epsilon_\tau}{8(n-1)}$ & 0 
       &$\frac{(n-4)\epsilon - 12n\epsilon_\tau}{8n(n-1)}$  & 0 \\
RR$_2$ &$\frac{\epsilon + 4\epsilon_\tau}{16(n-1)}$
       &$\frac{(4-n)\epsilon + 4(n+2)\epsilon_\tau}{8(n-1)}$ & 0 
       &$\frac{(n-4)\epsilon - 12n\epsilon_\tau}{8n(n-1)}$
       &$-\frac{(n-4)\epsilon - 12n\epsilon_\tau}{8n(n-1)}$ \\ 
\hline
\end{tabular}
\end{table}
\begin{table}
\caption{Fixed points with $\bar x^\ast \neq 0$}
\label{table:1a}
\begin{tabular}{rccccc} \hline
&$u^\ast$ &$\Delta^\ast$ &$\bar x^\ast$ &$z^\ast$ &$w^\ast$ \\ 
\hline
1  & 0 & 0 & $-\frac{\epsilon - 2\epsilon_\tau}{4n}$ & 0 
   &$\frac{\epsilon - 2\epsilon_\tau}{2n}$ \\
2  & 0 & 0 & $-\frac{\epsilon - 2\epsilon_\tau}{4n}$ & 0 & 0  \\
3  & 0 & 0 & $-\frac{\epsilon - 2\epsilon_\tau}{4n}$ 
   &$\frac{\epsilon - 2\epsilon_\tau}{2n}$ & 0 \\
4  & 0 & 0 & $-\frac{\epsilon - 2\epsilon_\tau}{4n}$ 
   &$\frac{\epsilon - 2\epsilon_\tau}{2n}$
   &$-\frac{\epsilon - 2\epsilon_\tau}{2n}$ \\
5  &$\frac{\epsilon - 2\epsilon_\tau}{4(n+8)}$ & 0 
   & $\frac{(n-4)(\epsilon - 2\epsilon_\tau)}{4n(n+8)}$ & 0 
   &$\frac{(\epsilon - 2\epsilon_\tau)(4-n)}{2n(n+8)}$ \\
6  &$\frac{\epsilon - 2\epsilon_\tau}{4(n+8)}$ & 0 
   &$\frac{(n-4)(\epsilon - 2\epsilon_\tau)}{4n(n+8)}$ & 0 & 0 \\
7  &$\frac{\epsilon - 2\epsilon_\tau}{4(n+8)}$ & 0 
   &$\frac{(n-4)(\epsilon - 2\epsilon_\tau)}{4n(n+8)}$ 
   &$\frac{(\epsilon - 2\epsilon_\tau)(4-n)}{2n(n+8)}$ & 0 \\
8  &$\frac{\epsilon - 2\epsilon_\tau}{4(n+8)}$ & 0 
   &$\frac{(n-4)(\epsilon - 2\epsilon_\tau)}{4n(n+8)}$ 
   &$\frac{(\epsilon - 2\epsilon_\tau)(4-n)}{2n(n+8)}$  
   &$-\frac{(\epsilon - 2\epsilon_\tau)(4-n)}{2n(n+8)}$  \\
9  & 0 & $\frac{-\epsilon + 4\epsilon_\tau}{8}$ 
   &$-\frac{\epsilon + 4\epsilon_\tau}{16n}$ & 0 
   &$\frac{3\epsilon - 4\epsilon_\tau}{8n}$ \\
10 & 0 & $\frac{-\epsilon + 4\epsilon_\tau}{8}$ 
   &$-\frac{\epsilon + 4\epsilon_\tau}{16n}$ & 0 & 0 \\
11 & 0 & $\frac{-\epsilon + 4\epsilon_\tau}{8}$ 
   &$-\frac{\epsilon + 4\epsilon_\tau}{16n}$ 
   &$\frac{3\epsilon - 4\epsilon_\tau}{8n}$ & 0 \\
12 & 0 & $\frac{-\epsilon + 4\epsilon_\tau}{8}$ 
   &$-\frac{\epsilon + 4\epsilon_\tau}{16n}$ 
   &$\frac{3\epsilon - 4\epsilon_\tau}{8n}$ 
   &$-\frac{3\epsilon - 4\epsilon_\tau}{8n}$ \\
13 & $\frac{\epsilon + 4\epsilon_\tau}{8(10-n)}$ 
   &$\frac{(n-4)\epsilon + 24\epsilon_\tau}{4(10-n)}$ 
   &$\frac{(n-4)(\epsilon + 4\epsilon_\tau)}{8n(10-n)}$ & 0 
   &$\frac{(4-n)3\epsilon -(n+8)4\epsilon_\tau}{4n(10-n)}$\\
14 &$\frac{\epsilon + 4\epsilon_\tau}{8(10-n)}$ 
   &$\frac{(n-4)\epsilon + 24\epsilon_\tau}{4(10-n)}$ 
   &$\frac{(n-4)(\epsilon + 4\epsilon_\tau)}{8n(10-n)}$ & 0 & 0 \\
15 &$\frac{\epsilon + 4\epsilon_\tau}{8(10-n)}$ 
   &$\frac{(n-4)\epsilon + 24\epsilon_\tau}{4(10-n)}$ 
   &$\frac{(n-4)(\epsilon + 4\epsilon_\tau)}{8n(10-n)}$ 
   &$\frac{(4-n)3\epsilon -(n+8)4\epsilon_\tau}{4n(10-n)}$ & 0 \\
16 &$\frac{\epsilon + 4\epsilon_\tau}{8(10-n)}$ 
   &$\frac{(n-4)\epsilon + 24\epsilon_\tau}{4(10-n)}$ 
   &$\frac{(n-4)(\epsilon + 4\epsilon_\tau)}{8n(10-n)}$ 
   &$\frac{(4-n)3\epsilon -(n+8)4\epsilon_\tau}{4n(10-n)}$ 
   &$-\frac{(4-n)3\epsilon -(n+8)4\epsilon_\tau}{4n(10-n)}$ \\
\hline 
\end{tabular}
\end{table}
As can be seen from Table\ \ref{table:1} we can separate all 
fixed points on groups consisting of two fixed points, for example 
group of Gaussian points (G) and group of renormalized Gaussian 
points (RG).
Each group has its proper set of critical exponents. 

Such classification of fixed points was used for description of phase 
transition in classical systems \cite{Chaves,Achiam2,LS} where for 
renormalized fixed points the critical exponents can be found according to 
Fisher \cite{Fisher} via critical exponents of appropriate 
non-renormalized values
\begin{equation}
\alpha_{\mathrm{renorm}}=-\frac{\alpha}{1-\alpha}, 
\quad \nu_{\mathrm{renorm}}=\frac{\nu}{1-\alpha}.
\label{frenorm}
\end{equation}

The eigenvalues $\lambda_i$ ($i = 1, \ldots, 5$) of flow equations 
linearized about the fixed point (\ref{u})-(\ref{w}) are indicated 
in Table\ \ref{table:2} and Table\ \ref{table:2a}.
\begin{table}
\caption{Eigenvalue of fixed points with $\bar x^\ast = 0$
($\lambda_{1,2}^R = \frac{1}{8(n-1)} \{-3n\epsilon + 4(n-4)\epsilon_\tau 
\pm [(5n-8)^2 \epsilon^2 + 24(16-12n-n^2)\epsilon\epsilon_\tau +
48(16-8n-5n^2)\epsilon_\tau^2 ]^{1/2} \}$)}
\label{table:2}
\begin{tabular}{lccccc} \hline
&$\lambda_1$ &$\lambda_2$ &$\lambda_3$ &$\lambda_4$ &$\lambda_5$\\ 
\hline
G$_1$  &$\epsilon-2\epsilon_\tau$ &$\epsilon$ &$\epsilon-2\epsilon_\tau$ 
       &$\epsilon-2\epsilon_\tau$ &$-\epsilon+2\epsilon_\tau$\\
G$_2$  &$\epsilon-2\epsilon_\tau$ &$\epsilon$ &$\epsilon-2\epsilon_\tau$ 
       &$\epsilon-2\epsilon_\tau$ &$\epsilon-2\epsilon_\tau$\\
RG$_1$ &$\epsilon-2\epsilon_\tau$ &$\epsilon$ &$\epsilon-2\epsilon_\tau$ 
       &$-\epsilon+2\epsilon_\tau$ &$-\epsilon+2\epsilon_\tau$\\
RG$_2$ &$\epsilon-2\epsilon_\tau$ &$\epsilon$ &$\epsilon-2\epsilon_\tau$
       &$-\epsilon+2\epsilon_\tau$ &$\epsilon-2\epsilon_\tau$ \\
H$_1$  &$-\epsilon+2\epsilon_\tau$ 
       &$\frac{(4-n)\epsilon+4(n+2)\epsilon_\tau}{n+8}$ 
       &$\frac{(4-n)(\epsilon-2\epsilon_\tau)}{n+8}$ 
       &$\frac{(4-n)(\epsilon-2\epsilon_\tau)}{n+8}$ 
       &$-\frac{(4-n)(\epsilon-2\epsilon_\tau)}{n+8}$\\
H$_2$  &$-\epsilon+2\epsilon_\tau$ 
       &$\frac{(4-n)\epsilon+4(n+2)\epsilon_\tau}{n+8}$ 
       &$\frac{(4-n)(\epsilon-2\epsilon_\tau)}{n+8}$ 
       &$\frac{(4-n)(\epsilon-2\epsilon_\tau)}{n+8}$ 
       &$\frac{(4-n)(\epsilon-2\epsilon_\tau)}{n+8}$\\
RH$_1$ &$-\epsilon+2\epsilon_\tau$ 
       &$\frac{(4-n)\epsilon+4(n+2)\epsilon_\tau}{n+8}$ 
       &$\frac{(4-n)(\epsilon-2\epsilon_\tau)}{n+8}$ 
       &$-\frac{(4-n)(\epsilon-2\epsilon_\tau)}{n+8}$ 
       &$-\frac{(4-n)(\epsilon-2\epsilon_\tau)}{n+8}$\\
RH$_2$ &$-\epsilon+2\epsilon_\tau$ 
       &$\frac{(4-n)\epsilon+4(n+2)\epsilon_\tau}{n+8}$ 
       &$\frac{(4-n)(\epsilon-2\epsilon_\tau)}{n+8}$ 
       &$-\frac{(4-n)(\epsilon-2\epsilon_\tau)}{n+8}$ 
       &$\frac{(4-n)(\epsilon-2\epsilon_\tau)}{n+8}$\\
U$_1$  &$-\frac12\epsilon-2\epsilon_\tau$ &$-\epsilon$ 
       &$-\frac12\epsilon-2\epsilon_\tau$
       &$\frac12\epsilon-2\epsilon_\tau$ 
       &$-\frac12\epsilon+2\epsilon_\tau$\\
U$_2$  &$-\frac12\epsilon-2\epsilon_\tau$ &$-\epsilon$ 
       &$-\frac12\epsilon-2\epsilon_\tau$
       &$\frac12\epsilon-2\epsilon_\tau$ 
       &$\frac12\epsilon-2\epsilon_\tau$\\
RU$_1$ &$-\frac12\epsilon-2\epsilon_\tau$ &$-\epsilon$ 
       &$-\frac12\epsilon-2\epsilon_\tau$
       &$-\frac12\epsilon+2\epsilon_\tau$ 
       &$-\frac12\epsilon+2\epsilon_\tau$\\ 
RU$_2$ &$-\frac12\epsilon-2\epsilon_\tau$ &$-\epsilon$ 
       &$-\frac12\epsilon-2\epsilon_\tau$
       &$-\frac12\epsilon+2\epsilon_\tau$ 
       &$\frac12\epsilon-2\epsilon_\tau$\\
R$_1$  &$\lambda_1^R$ &$\lambda_2^R$ 
       &$\frac{(4-n)(\epsilon+4\epsilon_\tau)}{4(n-1)}$
       &$\frac{(n-4)\epsilon-12n\epsilon_\tau}{4(n-1)}$ 
       &$-\frac{(n-4)\epsilon-12n\epsilon_\tau}{4(n-1)}$\\
R$_2$  &$\lambda_1^R$ &$\lambda_2^R$ 
       &$\frac{(4-n)(\epsilon+4\epsilon_\tau)}{4(n-1)}$
       &$\frac{(n-4)\epsilon-12n\epsilon_\tau}{4(n-1)}$ 
       &$\frac{(n-4)\epsilon-12n\epsilon_\tau}{4(n-1)}$\\
RR$_1$ &$\lambda_1^R$ &$\lambda_2^R$ 
       &$\frac{(4-n)(\epsilon+4\epsilon_\tau)}{4(n-1)}$
       &$-\frac{(n-4)\epsilon-12n\epsilon_\tau}{4(n-1)}$ 
       &$-\frac{(n-4)\epsilon-12n\epsilon_\tau}{4(n-1)}$\\
RR$_2$ &$\lambda_1^R$ &$\lambda_2^R$ 
       &$\frac{(4-n)(\epsilon+4\epsilon_\tau)}{4(n-1)}$
       &$-\frac{(n-4)\epsilon-12n\epsilon_\tau}{4(n-1)}$ 
       &$\frac{(n-4)\epsilon-12n\epsilon_\tau}{4(n-1)}$\\ 
\hline
\end{tabular}
\end{table}
\begin{table}
\caption{Eigenvalue of fixed points with $\bar x^\ast \neq 0$
($\lambda_{2,3}^{(9)} = \frac18 \{ 4\epsilon_\tau - 3\epsilon \pm  
[25 \epsilon^2 - 24 \epsilon \epsilon_\tau - 
240 \epsilon_{\tau}^2]^{1/2} \}$ and 
$\lambda_{1,2}^{(13)} = -\frac{1}{4(10-n)} \{ (16-n)\epsilon + 
4(n-4)\epsilon_\tau \pm [9(8-n)^2 \epsilon^2 + 
24(n^2 -4n-48)\epsilon \epsilon_\tau + 
16(n^2 +40n-464)\epsilon_\tau^2]^{1/2} \}$)}
\label{table:2a}
\begin{tabular}{rccccc} \hline
&$\lambda_1$ &$\lambda_2$ &$\lambda_3$ &$\lambda_4$ &$\lambda_5$\\ 
\hline
1  &$\epsilon-2\epsilon_\tau$ &$-\epsilon + 4\epsilon_\tau$ 
   &$-\epsilon + 2\epsilon_\tau$ 
   &$\epsilon-2\epsilon_\tau$ &$-\epsilon+2\epsilon_\tau$\\
2  &$\epsilon-2\epsilon_\tau$ &$-\epsilon + 4\epsilon_\tau$ 
   &$-\epsilon + 2\epsilon_\tau$ 
   &$\epsilon-2\epsilon_\tau$ &$\epsilon-2\epsilon_\tau$\\
3  &$\epsilon-2\epsilon_\tau$ &$-\epsilon + 4\epsilon_\tau$ 
   &$-\epsilon + 2\epsilon_\tau$ 
   &$-\epsilon+2\epsilon_\tau$ &$-\epsilon+2\epsilon_\tau$\\
4  &$\epsilon-2\epsilon_\tau$ &$-\epsilon + 4\epsilon_\tau$ 
   &$-\epsilon + 2\epsilon_\tau$
   &$-\epsilon+2\epsilon_\tau$ &$\epsilon-2\epsilon_\tau$ \\
5  &$-\epsilon+2\epsilon_\tau$ 
   &$\frac{(n-4)\epsilon + 24\epsilon_\tau}{n+8}$ 
   &$\frac{(n-4)(\epsilon - 2\epsilon_\tau)}{n+8}$ 
   &$\frac{(4-n)(\epsilon-2\epsilon_\tau)}{n+8}$ 
   &$-\frac{(4-n)(\epsilon-2\epsilon_\tau)}{n+8}$\\
6  &$-\epsilon+2\epsilon_\tau$ 
   &$\frac{(n-4)\epsilon + 24\epsilon_\tau}{n+8}$ 
   &$\frac{(n-4)(\epsilon - 2\epsilon_\tau)}{n+8}$ 
   &$\frac{(4-n)(\epsilon-2\epsilon_\tau)}{n+8}$ 
   &$\frac{(4-n)(\epsilon-2\epsilon_\tau)}{n+8}$\\
7  &$-\epsilon+2\epsilon_\tau$ 
   &$\frac{(n-4)\epsilon + 24\epsilon_\tau}{n+8}$ 
   &$\frac{(n-4)(\epsilon - 2\epsilon_\tau)}{n+8}$ 
   &$-\frac{(4-n)(\epsilon-2\epsilon_\tau)}{n+8}$ 
   &$-\frac{(4-n)(\epsilon-2\epsilon_\tau)}{n+8}$\\
8  &$-\epsilon+2\epsilon_\tau$ 
   &$\frac{(n-4)\epsilon + 24\epsilon_\tau}{n+8}$ 
   &$\frac{(n-4)(\epsilon - 2\epsilon_\tau)}{n+8}$ 
   &$-\frac{(4-n)(\epsilon-2\epsilon_\tau)}{n+8}$ 
   &$\frac{(4-n)(\epsilon-2\epsilon_\tau)}{n+8}$\\
9  &$\frac{\epsilon + 4\epsilon_\tau}{4}$ 
   &$\lambda_2^{(9)}$ &$\lambda_3^{(9)}$ 
   &$\frac{3\epsilon - 4\epsilon_\tau}{4}$ 
   &$-\frac{3\epsilon + 4\epsilon_\tau}{4}$\\
10 &$\frac{\epsilon + 4\epsilon_\tau}{4}$  
   &$\lambda_2^{(9)}$ &$\lambda_3^{(9)}$
   &$\frac{3\epsilon - 4\epsilon_\tau}{4}$ 
   &$\frac{3\epsilon - 4\epsilon_\tau}{4}$\\
11 &$\frac{\epsilon + 4\epsilon_\tau}{4}$  
   &$\lambda_2^{(9)}$ &$\lambda_3^{(9)}$
   &$-\frac{3\epsilon + 4\epsilon_\tau}{4}$ 
   &$-\frac{3\epsilon + 4\epsilon_\tau}{4}$\\ 
12 &$\frac{\epsilon + 4\epsilon_\tau}{4}$  
   &$\lambda^{(9)}_2$ &$\lambda^{(9)}_3$
   &$-\frac{3\epsilon + 4\epsilon_\tau}{4}$ 
   &$\frac{3\epsilon - 4\epsilon_\tau}{4}$\\
13 &$\lambda_1^{(13)}$ &$\lambda_2^{(13)}$ 
   &$\frac{(n-4)(\epsilon + 4\epsilon_\tau)}{2(10-n)}$
   &$\frac{(4-n)3\epsilon - (n+8)4\epsilon_\tau}{2(10-n)}$ 
   &$-\frac{(4-n)3\epsilon - (n+8)4\epsilon_\tau}{2(10-n)}$\\
14 &$\lambda_1^{(13)}$ &$\lambda_2^{(13)}$ 
   &$\frac{(n-4)(\epsilon + 4\epsilon_\tau)}{2(10-n)}$
   &$\frac{(4-n)3\epsilon - (n+8)4\epsilon_\tau}{2(10-n)}$ 
   &$\frac{(4-n)3\epsilon - (n+8)4\epsilon_\tau}{2(10-n)}$\\
15 &$\lambda_1^{(13)}$ &$\lambda_2^{(13)}$ 
   &$\frac{(n-4)(\epsilon + 4\epsilon_\tau)}{2(10-n)}$
   &$-\frac{(4-n)3\epsilon - (n+8)4\epsilon_\tau}{2(10-n)}$ 
   &$-\frac{(4-n)3\epsilon - (n+8)4\epsilon_\tau}{2(10-n)}$\\
16 &$\lambda_1^{(13)}$ &$\lambda_2^{(13)}$ 
   &$\frac{(n-4)(\epsilon + 4\epsilon_\tau)}{2(10-n)}$
   &$-\frac{(4-n)3\epsilon - (n+8)4\epsilon_\tau}{2(10-n)}$ 
   &$\frac{(4-n)3\epsilon - (n+8)4\epsilon_\tau}{2(10-n)}$\\ 
\hline
\end{tabular}
\end{table}

It should be noted that eigenvalues $\lambda_1$ and $\lambda_2$ 
for random (R) and renormalized random (RR) fixed points for 
$n > 1$ are complex.

Due to the relation
\begin{equation}
u = \lambda_s - \lambda_c^{(1)},
\label {condition}
\end{equation}
the fixed points with $\Delta^\ast = 0$ and $\bar x^\ast = 0$ 
align on parallel lines:
\begin{enumerate}
\item $\lambda_s = \lambda_c$ for the unconstrained system ($z=0$);
\item $\lambda_s = \lambda_c^{(1)}$ for the constrained 
system ($w = 0$); 
\item $\lambda_s - \lambda_c = (\epsilon - 2\epsilon_\tau) /4K_4(n + 8)$ 
for the unconstrained system ($z = 0$); 
\item $\lambda_s - \lambda_c^{(1)} = (\epsilon - 
2\epsilon_\tau) /4K_4(n + 8)$ for the constrained system ($w = 0$).
\end{enumerate}

It is easy to see that for the quantum phase transition
($\epsilon_\tau \neq 0$) in the case when the inequality  
$\epsilon < 2\epsilon_\tau$ is satisfied the fixed points lying 
on the two last lines are not situated in the stability range of the 
effective action (Fig.\ \ref{fig:1} and Fig.\ \ref{fig:2}). 
Thus the stability range for the quantum phase transition in the
constrained system remains such as in the mean-field theory 
in contrast with the classical phase transition where these 
fixed points are situated in the stability range and cause 
an effective reduction of this range \cite{Lyra,Skryabin97}.
However for the unconstrained system there is the unstable 
fixed point $G_1$ on the boundary of the stability range.
Its critical exponents coincide with critical exponents of the point $G_2$.
Analogous to the classical phase transition a separatrix going out 
this point divides the stability range in two regions. One is the region 
of accessibility of the fixed point $G_2$ and the other is
the region of phase space where flow trajectories approach to the 
``invariant'' line $G_2-G_1$.

\begin{figure}
\begin{center}
\resizebox{0.5\textwidth}{!}{\includegraphics{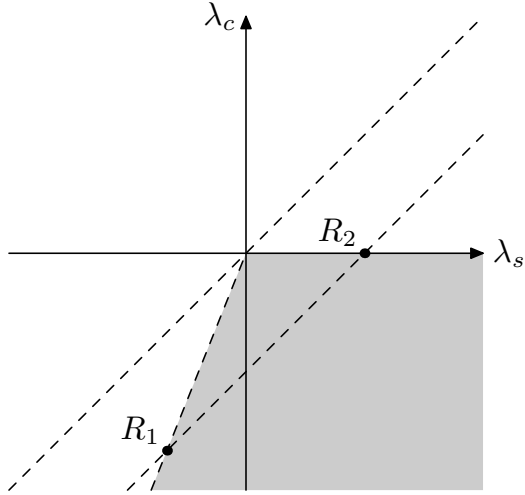}}
\end{center}
\caption{Phase diagram for the unconstrained system
($\Delta^\ast \neq 0, \quad \bar x^\ast = 0$)}
\label{fig:3}
\end{figure}

\begin{figure}
\begin{center}
\resizebox{0.5\textwidth}{!}{\includegraphics{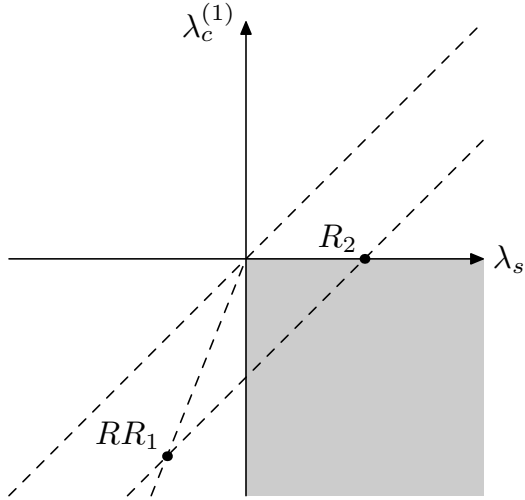}}
\end{center}
\caption{Phase diagram for the constrained system
($\Delta^\ast \neq 0, \quad \bar x^\ast = 0$)} 
\label{fig:4}
\end{figure}

Correspondingly, the fixed points with $\Delta^\ast \neq 0$ and 
$\bar x^\ast = 0$ are situated along parallel lines:
\begin{enumerate}
\item $\lambda_s - \lambda_c = (\epsilon + 4\epsilon_\tau) /16K_4(n-1)$ 
for the unconstrained system ($z = 0$); 
\item $\lambda_s -\lambda_c^{(1)} = 
(\epsilon + 4\epsilon_\tau) /16K_4(n-1)$ 
for the constrained system ($w = 0$).
\end{enumerate}
For the unconstrained system the fixed points lying on these lines for 
both the classical ($\epsilon_\tau = 0$) and the quantum phase transition 
are situated in the stability range (Fig.\ \ref{fig:3}). 
Thus the accessibility range of fixed points is also reduced 
in comparison with the mean-field result. 
For the constrained system not all points are in the range of stability 
(Fig.\ \ref{fig:4}).

The analysis of the renormalization group equations (\ref{u}), 
(\ref{x}) and (\ref{delta}) shows that as well as in the case 
of systems without hidden degrees of freedom 
\cite{Narayanan99a,Narayanan99b} 
the Gaussian fixed point G$_2$ is stable for $d > 4$ and unstable 
to disorder for $d < 4$.
The random fixed point R$_2$ is stable for 
$3n\epsilon > 4 (n-4) \epsilon_\tau$, $n > 4$ and 
$12n\epsilon_\tau > (n-4) \epsilon$.
It is easy to see that this point is always stable 
for $d < 4$ and $4 < n < n_c$ where $n_c =16$ 
for the particular case $\epsilon_\tau = \epsilon$.
It should be noted here that in the conventional theory this 
fixed point is stable for $d < 4$ and $n < 4$ 
\cite{Kirkpatrick96a,Skryabin99b}.
For $d < 4$ the stable unphysical fixed point U$_2$ is unaccessible 
from the physical range (according to definition the magnitude 
$\Delta$ should be positive) and for $d > 4$ it is unstable.
The Heisenberg fixed point H$_2$ for the pure quantum system 
becomes unstable because the dynamic critical exponent $z$ 
reduces the upper critical dimensionality.
There also are new fixed points with $\bar x^\ast \neq 0$.
As can be seen from Table\ \ref{table:2a} they are all unstable 
except the fixed point 14, which is stable for $n < 4$ and 
$3(4-n)\epsilon - 4(n+8)\epsilon_\tau < 0$. However, this fixed 
point is also unaccessible from physical range since the value of 
$\bar x^\ast$ is negative.   

Thus, we see that the quantum character of the phase transition in systems 
with hidden degrees of freedom with constraints leads to conclusion 
that for pure systems ($\Delta^\ast = \bar x^\ast = 0$) the phase 
transition is described by the Gaussian fixed point with 
``classical'' critical exponents of the mean-field theory whereas 
for the disordered systems 
($\Delta^\ast \neq 0,\quad \bar x^\ast \neq 0$) 
the rare regions destroy the phase transition for $n < 4$ similar 
to systems without hidden degrees of freedom  
\cite{Narayanan99a,Narayanan99b}.  
However, for $4 < n < n_c$ the phase transition is described by 
the random fixed point. 
Hidden degrees of freedom result in a decrease of the  
stability range with comparison to the result of the mean-field theory 
and existence of the range of the phase space in which the flow 
trajectories runaway from fixed points not intersecting the stability 
range of the mean-field.

\bibliography{data}

\begin{thebibliography}{10}

\bibitem{Griffiths}
R.~B. Griffiths, Phys. Rev. Lett. {\bf 23},  17  (1969).

\bibitem{Harris}
A.~B. Harris, J. Phys. C {\bf 7},  1671  (1974).

\bibitem{Dotsenko}
V. Dotsenko, A.~B. Harris, D. Sherrington, and R. Stinchcombe, J. Phys. A {\bf
  28},  3093  (1995).

\bibitem{Narayanan99a}
R. Narayanan, T. Vojta, D. Belitz, and T.~R. Kirkpatrick, Phys. Rev. Lett. {\bf
  82},  5132  (1999).

\bibitem{Narayanan99b}
R. Narayanan, T. Vojta, D. Belitz, and T.~R. Kirkpatrick, {\em On the critical
  behavior of disordered magnets: The relevance of rare regions},
  cond-mat/9905047.

\bibitem{Kirkpatrick96a}
T.~R. Kirkpatrick and D. Belitz, Phys. Rev. Lett. {\bf 76},  2571  (1996).

\bibitem{Kirkpatrick96b}
T.~R. Kirkpatrick and D. Belitz, Phys. Rev. B {\bf 53},  14364  (1996).

\bibitem{Skryabin99b}
Y.~N. Skryabin, A.~V. Chukin, and A.~V. Shchanov, Physica A {\bf 272},  162
  (1999).

\bibitem{Hertz}
J.~A. Hertz, Phys. Rev. B {\bf 14},  1165  (1976).

\bibitem{Chaves}
C.~M. Chaves, P. Lederer, and A.~A. Gomes, J. Phys. C {\bf 10},  3367  (1977).

\bibitem{Lyra}
M.~L. Lyra, M.~D. Coutinho-Filho, and A.~M. Nemirovsky, Phys. Rev. B {\bf 48},
  3755  (1993).

\bibitem{Skryabin97}
Y.~N. Skryabin and A.~V. Shchanov, Phys. Lett. A {\bf 234},  147  (1997).

\bibitem{Achiam1}
Y. Achiam and Y. Imry, Phys. Rev. B {\bf 12},  2768  (1975).

\bibitem{Achiam2}
Y. Achiam, J. Phys. C {\bf 10},  1491  (1977).

\bibitem{LS}
V.~M. Laptev and Y.~N. Skryabin, Phys. Stat. Sol. (b) {\bf 91},  K143  (1979).

\bibitem{Fisher}
M.~E. Fisher, Phys. Rev. B {\bf 176},  257  (1968).

\end{thebibliography}
\bibliographystyle{prsty}
\end{document}